\documentclass[12pt,preprint,graphicx]{aastex}

\DeclareGraphicsExtensions{.eps}

\begin{document}

\title{Discovery of a new radio galaxy within the error box of
the unidentified $\gamma$-ray source 3EG~J1735$-$1500}

\author{J.~A. Combi\altaffilmark{1}, G.~E. Romero\altaffilmark{1}, J.~M.
Paredes\altaffilmark{2}, D.~F. Torres\altaffilmark{3}, and M.
Rib\'o\altaffilmark{2} }


\altaffiltext{1} {Instituto Argentino de Radioastronom\'{\i}a,
C.C.5, (1894) Villa Elisa, Buenos Aires, Argentina}
\altaffiltext{2} {Departament d'Astronomia i Meteorologia,
Universitat de Barcelona, Av. Diagonal 647, 08028 Barcelona,
Spain} \altaffiltext{3} {Lawrence Livermore National Laboratory,
7000 East Ave., Livermore, CA 94550, USA }



\begin{abstract} We report the discovery of a new radio galaxy
within the location error box of the $\gamma$-ray source
3EG~J1735$-$1500. The galaxy is a double-sided jet source forming
a large angle with the line of sight. Optical observations reveal
a $V\sim18$ magnitude galaxy at the position of the radio core.
Although the association with the EGRET source is not confirmed at
the present stage, because there is a competing, alternative
$\gamma$-ray candidate within the location error contours which is
also studied here, the case deserves further attention. The new
radio galaxy can be used to test the recently proposed possibility
of $\gamma$-ray emitting radio galaxies beyond the already known
case of Centaurus~A. \end{abstract}


\section{Introduction} \label{introduction}

The quest for the identification of the $\gamma$-ray sources
detected by EGRET instrument of the Compton Gamma-Ray Observatory
is one of the most important challenges of high-energy
astrophysics in recent years. Among the 271 sources included in
the Third EGRET (3EG) catalog (Hartman et~al. 1999) there are a
few confirmed pulsars, 66 blazars, and a bunch of miscellaneous
identifications including one radio galaxy: the nearby
{Centaurus~A}, a Fanaroff-Riley (FR) Type~I active galaxy
(Sreekumar et~al. 1999).

Most identified blazars are strong flat-spectrum radio sources
(e.g. only 4 out 46 high-probability blazar detections in Mattox
et~al. (2001) analysis have fluxes below 1~Jy) with jets pointing
close to the line of sight. On the contrary, in the case of
{Centaurus~A} the viewing angle is rather large ($\sim70\degr$,
Bailey et~al. 1986). Very recently, Mukherjee et~al. (2002) have
suggested that the radio galaxy {NGC~6251} could also be
associated with an EGRET source. If this is confirmed it could
have important consequences because the spatial density of FR~I
radio galaxies is far higher than that of blazars, and hence,
despite that these objects are expected to be weaker $\gamma$-ray
emitters, there could be many other unidentified sources
associated with them.

In this paper we report the discovery of a new radio galaxy within
the location error box of the $\gamma$-ray source
 {3EG~J1735$-$1500}. The object was found during a
re-analysis of the main sources in the radio field around the
position of the EGRET detection. This research is part of a
systematic program to study potential low-energy counterparts of
unidentified $\gamma$-ray sources. Previous results regarding
compact radio sources were published by Torres et~al. (2001). Here
we present the results of a more detailed study and new
observations that led to the identification of the new radio
galaxy among the potential counterparts of
 {3EG~J1735$-$1500}. We also provide information on all
compact radio sources within the EGRET location error box,
including spectral index determinations when possible. We have
also found that the source  {PMN~J1738$-$1502} is probably a weak
flat-spectrum quasar with mild radio flux, which enhances its
probability of being the counterpart of  {3EG~J1735$-$1500}
respect to the a priori probability estimates by Mattox et~al.
(2001).

\begin{figure*}
\centering
\includegraphics[width=0.6\textwidth,height=10cm]{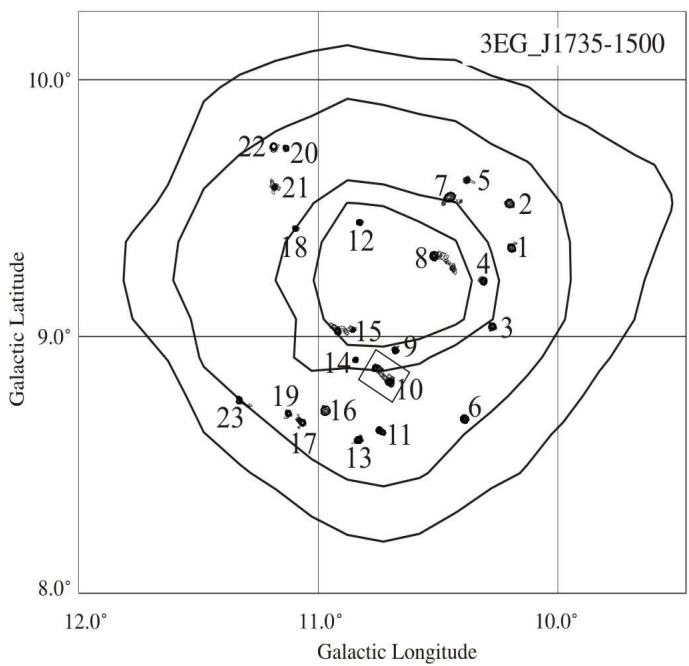}
\caption{$\gamma$-ray probability contours (50\%, 68\%, 95\% and
99\% from inside to outside) of the unidentified source
3EG~1735$-$1500, with the small-scale 1.4~GHz emission obtained
from the NVSS (where radio contours start at 1~mJy~beam$^{-1}$).
The position of the radio galaxy is marked with a rectangle.}
\label{fig1}
\end{figure*}

\section{Radio data analysis and results} \label{radio}

We have used the NRAO VLA Sky Survey (NVSS) (Condon et~al. 1998)
to study the small-scale radio emission within the inner 95~\%
location probability contours of the $\gamma$-ray source
{3EG~J1735$-$1500}. Twenty-three radio sources with flux density
$>10$~mJy at 1.4~GHz have been found, as can be seen in
Fig.~\ref{fig1}, where we have numbered the sources with
increasing galactic longitude. The main characteristics of these
sources are listed in Table~\ref{table1}. In particular, we
provide, from left to right, the identification number
corresponding to Fig.~\ref{fig1}, the galactic coordinates, flux
density at 1.4~GHz, spectral index (defined in such a way that
$S_{\nu}\propto\nu^{+\alpha}$) when observations at other
frequencies are available (365~MHz from Douglas et~al. 1996 and
4.8~GHz from Griffith \& Wright 1993) and the name in other
catalogs if any.

\begin{deluxetable}{lllll}
\label{table1} \tabletypesize{\scriptsize} \tablewidth{0pt}
\tablecaption{Radio sources within the 95\% $\gamma$-ray contour.}
 \tablehead{ \colhead{\#
} & \colhead{Coordinates}      & \colhead{$S_{1.4 \,\rm GHz}$  } &
\colhead{$\alpha$ } & \colhead{Other ID} \\
  \colhead{} & \colhead{$(l[\degr], b[\degr])$ } & \colhead{[mJy]}  &
\colhead{} & \colhead{}} \startdata

 1   & (10.22,+9.34)  & 71.00 &
$-1.34$  &   {TXS~1731$-$153}
\\ 2 & (10.23,+9.51)  &  128.96  &  $-1.2$   &  {TXS~1730$-$152}
\\ 3 & (10.29,+9.04)  &   32.06  & & \\ 4   &  (10.33,+9.21) &
31.68  & & \\ 5   & (10.39,+9.60)  &   10.88  & & \\ 6 &
(10.41,+8.69) & 177.57  &  $-1.04$  &  {TXS~1734$-$155} \\ 7   &
(10.46,+9.53)  & 133.77  & $-0.54$  &  {PMN~J1734$-$1502} \\ 8 &
(10.53,+9.31) & 169.72  & $-0.89$  &  {TXS~1732$-$150} \\
9   &  (10.69,+8.95)  &   24.74  &           & \\ 10  &
(10.74,+8.85)  &   55.64  &  $\la-0.9 $  & \\ 11  & (10.76,+8.65)
&   37.88  &           & \\ 12  &  (10.84,+9.43) &   17.07  & &
\\ 13  &  (10.84,+8.61)  &   46.00  & & \\ 14  & (10.86,+8.91) &
11.33  &           & \\ 15  & (10.94,+9.03)  & 28.62  & & \\ 16 &
(10.98,+8.72) &  330.40  & $-0.17$  & {PMN~J1738$-$1502} \\ 17 &
(11.08,+8.68)  &   19.15 & & \\ 18  &  (11.11,+9.42) &   32.44 & &
\\ 19  & (11.14,+8.71)  &   10.73  & & \\ 20  & (11.15,+9.72) &
9.97 &           & \\ 21  & (11.19,+9.58)  & 9.63  & & \\ 22  &
(11.19,+9.73) &   25.52  &           & \\ 23 & (11.34,+8.76)  &
38.43  & & \\
\enddata

\end{deluxetable}

\begin{figure*}
\centering
\includegraphics[width=0.4\textwidth,height=12cm]{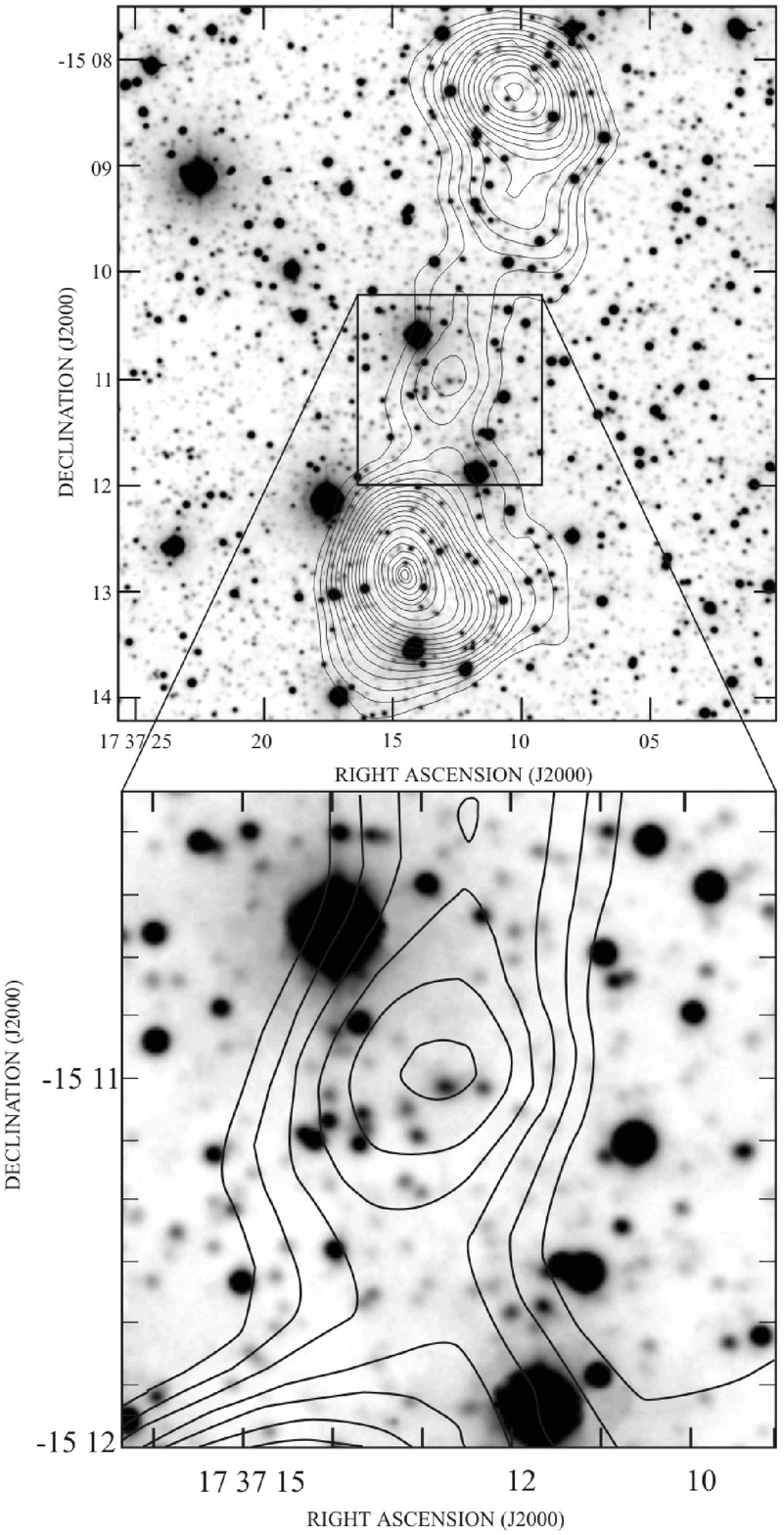}
\caption{{\bf Upper panel:} High-resolution radio image
of the galaxy
 {J1737$-$15} at 1.4~GHz overlapped to the optical image
obtained with the Calar Alto 2.2-m telescope using a Johnson's $I$
filter. Radio contours are shown in steps of 0.6~mJy~beam$^{-1}$,
starting from 1~mJy~beam$^{-1}$. {\bf Lower panel}: An enlargement
of the central region of the radio galaxy showing the possible
host galaxy. Radio contours are in steps of 0.3~mJy~beam$^{-1}$,
starting from 1.3~mJy~beam$^{-1}$.} \label{fig2}
\end{figure*}

With the exceptions of sources no.~8 and 10, all entries in
Table~\ref{table1} correspond to compact objects, at least at the
present angular resolution (43\arcsec). Source no.~8, which
appears as extended in our map, is actually a composite of three
different (and probably compact) weak sources listed as no.~7, 8,
and 9 in Torres et~al. (2001). Source no.~10, instead, is really
extended. We shall name this source hereafter as
 {J1737$-$15}, according to its radio coordinates. Its
morphology at 1.4~GHz (Fig.~\ref{fig2}, upper panel) is typical of
double-sided FR II radio galaxies. There is a nuclear component as
well as two jets ending in large radio lobes. The angular
dimensions of  {J1737$-$15} are $\sim0.04\degr\times0.11\degr$,
with an integrated flux of $55.6\pm1.5$~mJy at 1.4~GHz. The flux
density of the northern component is $S_{\rm
north}=21.6\pm0.5$~mJy, whereas the southern one is $S_{\rm
south}=30.7\pm1.3$~mJy. The central component has a flux density
of $S_{\rm core}=3.80\pm 0.05$~mJy, and an estimated J2000.0 ICRS
position of $\alpha=17^{\rm h}~37^{\rm m}~12.9^{\rm s}\pm0.3^{\rm
s}$, $\delta=-15\degr~11\arcmin~02\arcsec\pm15\arcsec$.

The source is clearly non-thermal since it is not detected at
4.8~GHz (Griffith \& Wright 1993). From the 4.8-GHz survey
sensitivity we can infer an average steep spectral index
$\alpha\la-0.9$, but it is not possible to say, within the present
resolution, what is the index at the core because of contamination
from the radio lobes, which have surely steeper spectra.
The general morphology clearly supports the hypothesis of a radio
galaxy with jets nearly perpendicular to the line of sight. Should
the association between the radio galaxy and the EGRET source be
confirmed,  {J1737$-$15} would be the second EGRET source (apart
 {Centaurus~A}) to be detected with a large inclination
angle.

\section{Optical observations} \label{optical}

With the aim of finding the optical counterpart of the radio
galaxy we have made $VR\,I$ deep photometric observations of
 {J1737$-$15} on 2002 June 9 at Calar Alto (Almer\'{\i}a,
Spain) with the 2.2-m telescope of the Centro Astron\'omico
Hispano-Alem\'an (CAHA). Images were obtained using the
Ritchey-Chr\'etien focus and CAFOS, with a scale factor of
$0.53\arcsec$ per pixel and a $8.8\arcmin\times8.8\arcmin$ field
of view. A series of 600~s images of our target, as well as 200~s
images of standard stars from the SA104 field of Landolt (1992),
were obtained through the $V$, $R$ and $I$ Johnson filters. The
observations were reduced using standard procedures (bias and dark
subtraction and flat-field correction) within the IRAF software
package. We performed a detailed astrometric reduction of the
images using 20 field stars present in the USNO-A2.0 catalog
(Monet et~al. 1999), with an estimated astrometric accuracy of
$0.3\arcsec$.

We show in Fig.~\ref{fig2} the obtained image through the $I$
Johnson filter, together with the contours of the radio source
from the NVSS. It is clear at first sight the presence of a
slightly elongated (east-west) optical object close to the center
of the radio source. The fitted J2000.0 ICRS position of this
object is $\alpha=17^{\rm h}~37^{\rm m}~12.744^{\rm
s}\pm0.021^{\rm s}$,
$\delta=-15\degr~11\arcmin~01.14\arcsec\pm0.30\arcsec$, well
within the previously obtained error box in position for the core
of the radio source. We also obtained absolute photometry of this
object, which is believed to be accurate only to $\pm$0.2
magnitudes, because the crowded field around our target prevented
a good estimate of the background. The obtained magnitudes are
$V=18.3\pm0.2$, $R=16.9\pm0.2$, and $I=15.1\pm0.2$. Using the
galactic extinction estimates from Schlegel et~al. (1998),
implemented within the NASA/IPAC Extragalactic Database (NED), we
obtain a color excess of $E(B-V)=0.495$ in the direction of
{J1737$-$15}. However, as pointed out by Arce \& Goodman (1999)
the values provided by the Schlegel's model could be overestimated
by a factor 1.3--1.5 in regions of smooth extinction with
$E(B-V)>0.15$~mag, as it happens in our case. As a resonable
value, we have considered that the NED value is overestimated by a
factor $1.3\pm0.2$, and used an extinction of
$E(B-V)=0.38\pm0.05$. This provides the following extinctions:
$A_V=3.315~E(B-V)=1.3\pm0.2$, $A_R=2.673~E(B-V)=1.0\pm0.1$ and
$A_I=1.940~E(B-V)=0.7\pm0.1$ (Schlegel et~al. 1998). Using these
values we obtain the following dereddened magnitudes:
$V=17.0\pm0.3$, $R=15.9\pm0.2$, and $I=14.4\pm0.2$. These values
imply color indices of $V-R=1.1\pm0.3$ and $R-I=1.5\pm0.3$ (where
the errors have been computed directly from the uncertainties in
measured magnitudes and assumed $E(B-V)$).

If we suppose that this optical object is a star, the ranges of
obtained dereddened colors only allow possibilities such as M3III
or M0I, according to Ducati et~al. (2001). However, comparison of
the obtained apparent dereddened visual magnitude with the
expected absolute visual magnitude (Wainscoat et~al. 1992) implies
distances of 33 and 465~kpc, respectively. The second value is
unrealistic, while the first one implies a galactic height
($b=9.2\degr$) of $\sim5$~kpc, in clear disagreement of what is
typically found for M3III stars. Therefore, the dereddened colors
are clearly incompatible with the object being a single galactic
star located by chance in the same direction of the radio galaxy
core.

On the other hand, since the source is elongated, a 2-D Gaussian
fit can provide information about the ellipticity and position
angle. We have performed several Gaussian fits using different
fitting box sizes. In all cases the ellipticity of the source
seems to be around 0.14, while the position angle changes from
$85\degr$ (counterclockwise from north) to $75\degr$ when reducing
the size of the fitting box. In any case, the object is extended
practically in the east-west direction, i.e., perpendicular to the
radio jets. All these facts suggest that our identified optical
counterpart is a galaxy, at the center of which the jets visible
at radio wavelengths are produced.

In addition to the optical counterpart we have looked for infrared
and X-ray sources at the position of  {J1737$-$15}. Filtered IRAS
images (Wheelock et~al. 1991) at 12 and 100 microns of the region
show no particular infrared enhancement at the central position of
the radio galaxy. The ROSAT All-Sky Survey (0.1--2.4~keV) shows
three X-ray sources inside of the $\gamma$-ray 95\% contour but
none of them is close to
 {J1737$-$15}.

\begin{figure*}
\centering
\includegraphics[width=0.6\textwidth,height=10cm]{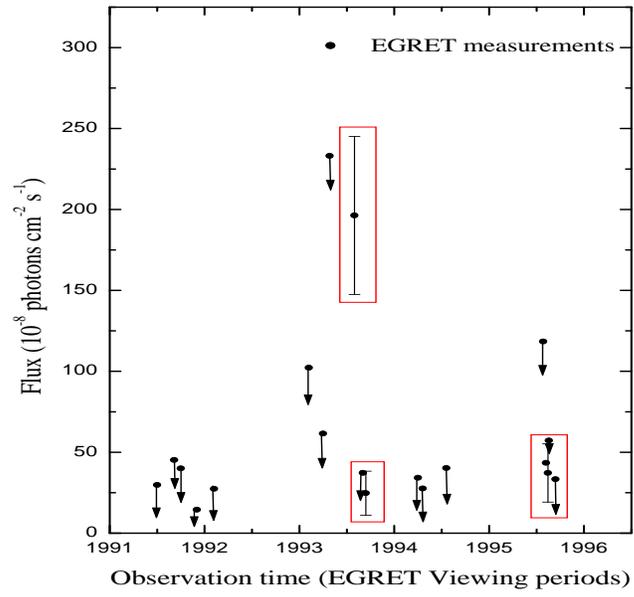} \caption{The $\gamma$-ray history of
{3EG~J1735$-$1500}. Boxes indicate source detections as opposed to
upper limits.} \label{fig3}
\end{figure*}

\section{Discussion} \label{discussion}

Depending on the jet and ambient medium parameters, most
double-sided radio sources have sizes below $\sim300$~kpc
(Begelman et~al. 1984). In the case of the radio galaxy reported
here, and using standard Friedmann-Robertson-Walker formulae, this
size translates into a possible distance smaller than 350~Mpc. If
 {3EG~J1735$-$1500} is indeed the result of $\gamma$-ray
emission in  {J1737$-$15}, the intrinsic luminosity at
$E>100$~MeV, at the distance quoted, should be smaller than
$2\times10^{44}$~erg~s$^{-1}$, although mild beaming could reduce
this figure. A correct determination of the distance requires a
knowledge of the redshift that is still unknown. The information
we provide here for the optical counterpart will be useful to
guide future spectroscopic observations. As in the case of
 {Centaurus~A} and  {3EG~J1621+8203}, the luminosity
needed to generate the $\gamma$-ray source observed is several
orders of magnitude less than those required for typical EGRET
blazars ($10^{45}$--$10^{48}$~erg~s$^{-1}$).

The $\gamma$-ray variability history of  {3EG~J1735$-$1500} is
shown in Fig.~\ref{fig3}. The source was detected only in three
EGRET viewing periods (VPs) --in the remaining VPs only upper
limits to the flux were established-- and appears to be variable.
Torres et al.'s (2001) index $I$ gives $I=8.86$, which place this
source as one of the most variable gamma-ray sources in the EGRET
catalog. Tompkins' (1999) index $\tau$ is $\tau=1.09_{     0.00}^{
10.1}$, which, although inconclusive due to its lower limit, the
central value and upper limit of the index also place this source
as one of the most variable ones in the Catalog. (For a detailed
discussion on variability indices see Torres, Pessah and Romero
2001).  The likely variability status of this source might imply
that the $\gamma$-ray emission is produced in the central region
and not in the extended lobes, if related with the discovered
radio galaxy. The photon spectral index seems to be unusually
steep with a value $\Gamma=3.24\pm0.47$, steeper than the average
of the $\gamma$-ray blazars spectra, a property also shared by
{Centaurus~A} and  {3EG~J1621+8203}.

Since  {3EG~J1735$-$1500} is at a low galactic latitude it could
be possible that some galactic object not visible at radio
wavelengths be responsible of the high-energy emission. However,
the source is not coincident with early-type stars, OB
associations, or galactic X-ray binaries (Romero et~al. 1999).

The only other known potential counterpart in addition to the
galaxy discussed here is the source  {PMN~J1738$-$1502}. It is a
compact radio source with a total flux at 1.4 MHz of $\sim0.3$ Jy
(source no.~16 in Table~\ref{table1}). We have calculated for this
object a flat spectral index $\alpha=-0.17$, which makes of it a
clear blazar candidate. Mattox et~al. (2001) give an a priori
probability of only 0.07 \% for a physical association. The a
posteriori probability (see Mattox et al.'s paper for details) is
slightly higher but still very small: 0.35 \%. The radio source is
weaker than most high-probability $\gamma$-ray blazars, but
similar to most plausible (i.e. with lower a priori probability)
identifications. Consequently, this source should be considered a
serious alternative counterpart of the EGRET source despite the
low probabilities given by Mattox et~al. (2001). The object is
similar to the plausible identification proposed recently by
Wallace et~al. (2002) in the case of
 {3EG~J2006$-$2321}, although we notice that in the case of
 {3EG~J1735$-$1500} the high-energy photon index is far
softer ($\Gamma=3.24\pm0.47$ vs $\Gamma=2.47\pm0.44$). Since
$\gamma$-ray blazars tend to be highly polarized and variable, we
are planning optical polarization observations of this source in
order to confirm its blazar nature.


One of the main problems to explain the synchrotron emission in
radio galaxies is that this radiation is distributed rather
uniformly along distances $\sim100$~kpc. To explain how the
continuous acceleration of electrons take place in these huge
sources remains as one of the most difficult topics in jet theory.
Recently, it has been suggested by Neronov et~al. (2002) that in
some cases the relativistic leptonic population can be locally
created by very-high energy $\gamma$-rays produced at the central
source and injected into the jet, where they produce pairs through
interactions with the 2.7~K cosmic background radiation. In this
scenario lower energy (GeV) photons would escape from the radio
galaxy and could even traverse the diffuse infrared background
without further interaction reaching the Earth. The $\gamma$-rays
could be produced in the central engine by photon-pion process
involving disk photons (Neronov et~al. 2002). Future
multifrequency observations of {J1737$-$15} could help to test the
proposal of an association with  {3EG~J1735$-$1500}. If the
high-energy emission is confirmed, this radio galaxy could become
an important natural laboratory to test the theories of
non-thermal emission in extragalactic jet sources.

\section{Conclusions}

We have discovered a new radio galaxy inside the error box of the
$\gamma$-ray source  {3EG~J1735$-$1500}. It is a double-sided FR
II type of object. Its linear angular size implies that it is
relatively nearby. We have also identified the host optical
galaxy, which appears elongated in a direction perpendicular to
the radio structure. Future spectroscopic observations will help
to fix the distance to this object. If it turns out to be closer
that 100~Mpc, the radio galaxy could be a potential acceleration
site of ultra-high energy cosmic rays, avoiding the so-called
Greisen-Zatsepin-Kuz'min (GZK) cut-off (see Rachen \& Biermann
1993).

We have also study the other important likely potential
counterpart of  {3EG~J1735$-$1500}. We have found that it is a
flat-spectrum compact radio source, possibly a blazar, and hence
perhaps a $\gamma$-ray emitting object. The remaining sources
within the location confidence contours of the EGRET detection
seem to be uninteresting from the point of view of the high-energy
emission. Forthcoming observational studies of these two objects
will shed light on the real nature of  {3EG~J1735$-$1500}.

\begin{acknowledgements}

This work was partially supported by the Argentinian agencies
CONICET (PIP 0430/98) and ANPCT (PICT 03-04881), as well as by the
DGI of the Ministerio de Ciencia y Tecnolog\'{\i}a (Spain) under
grant AYA2001-3092 and the European Regional Development Fund
(ERDF/FEDER). Additional support was provided by Fundaci\'on
Antorchas. The work of D.F.T. was performed under the auspices of
the U.S. Department of Energy, National Nuclear Security
Administration, by University of California Lawrence Livermore
National Laboratory under contract No. W-7405-Eng-48. D.F.T. is
Lawrence Fellow in Astrophysics. M.~R. is supported by a
fellowship from CIRIT (Generalitat de Catalunya, ref.
1999~FI~00199). We thank the German- Spanish Astronomical Centre,
Calar Alto, operated by the Max-Planck-Institut f\"ur Astronomie
(Heidelberg) jointly with the Spanish National Commission for
Astronomy, for carrying out the observations under the Director's
Discretionary Time Service.

\end{acknowledgements}

\end{document}